\documentclass[a4paper,11pt]{article}
\usepackage{pos}
\usepackage{scalerel}

\newcommand{\Eq}[1]{Eq.~(\ref{#1})}

\newcommand{\Fig}[1]{Fig.~\ref{#1}}

\title{Gauge Dependence of $c\bar{c}$ Potential from Nambu-Bethe-Salpeter Wave Function in Lattice QCD}
\ShortTitle{Gauge Dependence of $c\bar{c}$ Potential from NBS Wave Function in LQCD}

\author*{Tianchen Zhang}
\author{Noriyoshi Ishii}

\affiliation{Research Center for Nuclear Physics (RCNP), Osaka University,\\
10-1 Mihoga-oka, Ibaraki-shi, Osaka 567-0047, Japan}

\emailAdd{zhangtc@rcnp.osaka-u.ac.jp}

\abstract{We study the gauge dependence of $c\bar{c}$ potentials extracted from Nambu-Bethe-Salpeter (NBS) wave functions.
The potentials are obtained using the HAL QCD potential method, with an extension introduced by Kawanai and Sasaki for self-consistent determination of the charm quark mass.
A systematic comparison is conducted between results obtained in the Coulomb and Landau gauges.
The numerical calculations are performed using 2+1 flavor QCD gauge configurations with the charm quark treated in the quenched approximation.
We find that the central potentials in both gauges show excellent agreement at short distances but exhibit discrepancies at large distances.
We attribute these discrepancies to insufficient suppression of excited-state contamination in the Landau gauge, which aﬀects the linear-rising behavior of the potential at large distance.}

\FullConference{The 41st International Symposium on Lattice Field Theory (LATTICE2024)\\
 28 July - 3 August 2024\\
Liverpool, UK\\}

\begin{document}
\maketitle

\section{Motivation}
This work presents a detailed investigation of the gauge dependence of $c\bar{c}$ potentials and charm quark masses determined via the NBS amplitude method, which originated from the HAL QCD method \cite{PhysRevLett.99.022001,10.1143/PTP.123.89,ISHII2012437}.
While the gauge dependence of these quantities is intrinsically important, previous studies have predominantly employed the Coulomb gauge \cite{PhysRevLett.107.091601,PhysRevD.85.091503,PhysRevD.89.054507,PhysRevD.92.094503,PhysRevD.94.114514,Watanabe2021,PhysRevD.105.074510,Ikeda:2011mW,Iida:20123l,10.1143/PTP.128.941}, with one notable exception in an unpublished work by Hideaki Iida.
Their study, conducted in the Landau gauge, suggests that this gauge choice may influence the long-distance behavior of the central potential, though detailed results remain unavailable.
Such behavior contradicts the expectation that the $c\bar{c}$ potentials obtained from the NBS amplitude method in both Coulomb and Landau gauges should converge to the unique static $q\bar{q}$ potential obtained from the Wilson loop in the limit $m_c \rightarrow \infty$.

Furthermore, the present study extends to investigating diquark properties within quark-diquark systems, particularly focusing on diquark mass determination.
The $c\bar{c}$ system serves as an ideal simplified model for understanding the more complex quark-diquark system, as it shares fundamental characteristics with the diquark while maintaining analytical tractability.
This approach allows for the development of necessary formalism and numerical methods before addressing the full complexity of the $q\bar{q}$ system.

\section{Formalism}
We review the formalism of the Nambu-Bethe-Salpeter (NBS) amplitude approach, i.e., the extension of the HAL QCD potential method by Kawanai and Sasaki for calculating the $c\bar{c}$ potential with charm quark mass determined self-consistently \cite{PhysRevLett.107.091601,PhysRevD.85.091503,PhysRevD.89.054507,PhysRevD.92.094503,PhysRevD.94.114514,10.1143/PTP.128.941}.

The equal-time NBS amplitude of the charmonium system in the center-of-momentum (c.m.) frame, under a specific gauge-fixing condition, is defined as
\begin{equation}\label{eq:NBS_def}
  \phi_{\Gamma, n}(\mathbf{r}\equiv\mathbf{y} - \mathbf{x}) \equiv \scaleobj{0.85}{\frac{1}{V}\sum_{\mathbf{\Delta}}\sum_{c}}\langle 0|\bar{q}_c(\mathbf{x}+\mathbf{\Delta})\Gamma q_c(\mathbf{y}+\mathbf{\Delta})|\Gamma, n\rangle.
\end{equation}
Here, $V$ denotes the spatial volume of the system in a finite box, with $\sum_{\mathbf{\Delta}}$ corresponding to the projection onto zero total momentum states.
$q_c(x)$ denotes the Dirac field for the charm quark with color index $c=1,2,3$, and $\Gamma$ represents a Dirac $\gamma$ matrix.
For this study, we choose $\Gamma=\gamma_5$ for the pseudoscalar (PS) state ($J^P = 0^-$) and $\Gamma=\gamma_i~(i=1,2,3)$ for the vector (V) state ($J^P=1^-$).
The state $|\Gamma,n\rangle$, defined in the c.m. frame, denotes the $n$-th excited eigenstate of the $c\bar{c}$ system in the $\Gamma$ channel.
By setting $\mathbf{x}=\mathbf{0}$ without loss of generality, \Eq{eq:NBS_def} simplifies to
\begin{equation}
  \phi_{\Gamma, n}(\mathbf{r}) = \scaleobj{0.85}{\frac{1}{V}\sum_{\mathbf{\Delta}}\sum_{c}}\langle 0|\bar{q}_c(\mathbf{\Delta})\Gamma q_c(\mathbf{r}+\mathbf{\Delta})|\Gamma, n\rangle
\end{equation}
This defines the \textit{NBS wave function} of the $c\bar{c}$ system.

The NBS wave functions can be extracted from four-point correlators of charm quark fields, defined as
\begin{equation}\label{eq:4pt}
  C_\Gamma(\mathbf{r},t;t_{\rm src})\equiv\scaleobj{0.85}{\frac{1}{V}\sum_{\mathbf{\Delta}}\sum_{c}}\bigl\langle 0\bigl|T\bigl[\bar{q}_c(\mathbf{\Delta}, t)\Gamma q_c(\mathbf{\Delta}+\mathbf{r}, t)\mathcal{J}_\Gamma^\dagger(t_{\rm src})\bigr]\bigr|0\bigr\rangle,
\end{equation}
where $T$ denotes time ordering, and the wall source operator for $c\bar{c}$ at certain time $t$ is given as
\begin{equation}
  \mathcal{J}_\Gamma(t) \equiv \scaleobj{0.85}{\sum_{\mathbf{x}', \mathbf{y}'}\sum_{c'}}\bar{q}_{c'}(\mathbf{x'}, t)\Gamma q_{c'}(\mathbf{y'}, t).
\end{equation}

We set the source time to zero without loss of generality and drop the explicit notation of $t_{\rm src}$, writing $C_\Gamma(\mathbf{r},t)$ instead of $C_\Gamma(\mathbf{r},t;t_{\rm src})$.
By inserting the complete set $\mathbb{I}=\sum_{n}|\Gamma,n\rangle\langle \Gamma, n|$, the correlator can be expressed as a linear combination of NBS wave functions:
\begin{equation}\label{eq:4pt->NBS}
  C_\Gamma(\mathbf{r}, t) = \scaleobj{0.85}{\sum_{n}} a_{\Gamma,n} \phi_{\Gamma,n}(\mathbf{r}) e^{-E_{\Gamma,n}t},
\end{equation}
where $a_{\Gamma,n} \equiv \langle \Gamma,n|\mathcal{J}_\Gamma^\dagger(0)|0 \rangle$ and $E_{\Gamma,n}$ denotes the relativistic energy of the state $|\Gamma,n\rangle$.
Contributions from excited states are exponentially suppressed in the large-$t$ limit.
For the ground-state charmonium in the $\Gamma$ channel, The energy eigenvalue $E_{\Gamma,0}$ can be replaced by the rest mass $M_\Gamma$ due to the system being in the c.m. frame.
In what follows, we consider the \textit{ground-state NBS wave function} and adopt a simplified notation $\phi_\Gamma(\mathbf{r}) \equiv \phi_{\Gamma,0}(\mathbf{r})$ for brevity.

The S-wave projection of the ground-state NBS wave function is approximated by $\phi_{\Gamma}(\mathbf{r})\rightarrow\frac{1}{48}\sum_{g\in O_h}\phi_{\Gamma}(g^{-1}\mathbf{r})$ on the lattice, corresponding to the $A_1^+$ irreducible representation of the cubic symmetry group $O_h$.
While this projection contains higher partial-wave ($l\geq4$) contributions, such contamination becomes negligible for compact bound states when the lattice spacing is sufficiently fine and the spatial volume is large.

The NBS wave functions are assumed to satisfy the stationary Schrödinger equation \cite{10.1143/PTP.123.89}
\begin{equation}\label{eq:Seq_for_ccbar}
  E_\Gamma\phi_\Gamma(\mathbf{r}) = \bigl(\hat{K} + \hat{V}\bigr)\phi_\Gamma(\mathbf{r}),
\end{equation}
where $E_\Gamma \equiv M_\Gamma - 2m_c$ represents the "binding energy" of the charmonium system.
The rest mass $M_\Gamma$ is determined from the two-point correlators, while $m_c$ remains to be determined through the Kawanai-Sasaki condition \cite{PhysRevLett.107.091601}.
The kinetic energy operator is defined as $\hat{K} \equiv -\frac{\nabla^2}{2\mu}$ with the reduced mass $\mu = m_c/2$, and $\hat{V}$ denotes the potential operator.
The Laplacian operator $\nabla^2$ is discretized on the lattice using the nearest-neighbor finite-difference scheme as $\nabla^2\phi(\mathbf{x}) \simeq \sum_{i=x,y,z}\big[\phi(\mathbf{x}+a\mathbf{e}_i) + \phi(\mathbf{x}-a\mathbf{e}_i) - 2\phi(\mathbf{x})\big]/a^2$, where $a$ denotes the lattice spacing, and $\mathbf{e}_i$ represents the unit vector in the $i$-th spatial direction with $i=x,y,z$.

In the leading order of the derivative expansion, the potential operator $\hat{V}$ takes the form
\begin{equation}\label{eq:Vhat_expand}
  \hat{V} \simeq V_0(\mathbf{r}) + V_{\rm S}(\mathbf{r}) \mathbf{s}_1 \cdot \mathbf{s}_2 + O(\nabla),
\end{equation}
where $V_0(\mathbf{r})$ is the (spin-independent) central potential, and $V_{\rm S}(\mathbf{r})$ denotes the spin-dependent potential for the spin-spin interaction.
$\mathbf{s}_1$ and $\mathbf{s}_2$ represent the spin operators for the anti-charm quark ($\bar{c}$) and the charm quark ($c$), respectively.
The tensor potential vanishes under the S-wave projection.

The spin operator $\mathbf{s}_1 \cdot \mathbf{s}_2$ in the PS and V channels can be substituted by its eigenvalues $-3/4$ and $1/4$, respectively
Consequently, \Eq{eq:Seq_for_ccbar} simplifies to
\begin{equation}\label{eq:Seq_simplified}
  \begin{gathered}
    E_{\rm PS}\phi_{\rm PS}(\mathbf{r}) = \biggl(-\frac{\nabla^2}{2\mu} + V_0(\mathbf{r}) - \frac{3}{4}V_{\rm S}(\mathbf{r}) \biggr)\phi_{\rm PS}(\mathbf{r}), \\
    E_{\rm V}\phi_{\rm V}(\mathbf{r}) = \biggl(-\frac{\nabla^2}{2\mu} + V_0(\mathbf{r}) + \frac{1}{4}V_{\rm S}(\mathbf{r}) \biggr)\phi_{\rm V}(\mathbf{r}).
  \end{gathered}
\end{equation}
From the above equations, we can derive explicit expressions for $V_0(\mathbf{r})$ and $V_{\rm S}(\mathbf{r})$:
\begin{gather}
  V_0(\mathbf{r}) = \frac{1}{4 m_c}\Bigl[3\frac{\nabla^2 \phi_{\rm V}(\mathbf{r})}{\phi_{\rm V}(\mathbf{r})} + \frac{\nabla^2 \phi_{\rm PS}(\mathbf{r})}{\phi_{\rm PS}(\mathbf{r})}\Bigr] + E_{\rm ave}, \label{eq:spin_indep_pot}\\
  V_{\rm S}(\mathbf{r}) = \frac{1}{m_c}\Bigl[\frac{\nabla^2 \phi_{\rm V}(\mathbf{r})}{\phi_{\rm V}(\mathbf{r})} - \frac{\nabla^2 \phi_{\rm PS}(\mathbf{r})}{\phi_{\rm PS}(\mathbf{r})}\Bigr]+\Delta E, \label{eq:spin_dep_pot}
\end{gather}
where $E_{\rm ave}$ is defined as $\frac{1}{4}(3M_{\rm V} + M_{\rm PS})-2m_c$, and $\Delta E \equiv M_{\rm V} - M_{\rm PS}$ representing the hyperfine mass splitting between the ground states of the PS and V channels.

In the quark model, the spin-dependent potential $V_{\rm S}(\mathbf{r})$ originates from the color-magnetic interaction (spin-color interaction).
This potential is proportional to the delta function, indicating that $V_{\rm S}(\mathbf{r})$ is characterized by an extremely short-range behavior.
Based on this observation, Kawanai and Sasaki proposed a condition, known as the \textit{Kawanai-Sasaki condition}:
\begin{equation}\label{eq:KS_condition_for_Vs}
  \lim_{r\rightarrow \infty} V_{\rm S}(\mathbf{r}) = 0.
\end{equation}
By combining \Eq{eq:KS_condition_for_Vs} with \Eq{eq:spin_dep_pot}, we obtain
\begin{equation}\label{eq:KS_condition}
  m_c = -\lim_{r\rightarrow \infty}F_{\rm KS}(\mathbf{r}),
\end{equation}
where we introduce the \textit{Kawanai-Sasaki function} $F_{\rm KS}(\mathbf{r})$ defined by
\begin{equation}\label{eq:KS_function}
  F_{\rm KS}(\mathbf{r}) = \frac{1}{M_{\rm V} - M_{\rm PS}} \biggl[\frac{\nabla^2 \phi_{\rm V}(\mathbf{r})}{\phi_{\rm V}(\mathbf{r})} - \frac{\nabla^2 \phi_{\rm PS}(\mathbf{r})}{\phi_{\rm PS}(\mathbf{r})}\biggr].
\end{equation}

\section{Numerical results}
\subsection{Lattice Setup}
We use the $2+1$ flavor QCD gauge configurations on an $L^3 \times T = 32^3 \times 64$ lattice generated by PACS-CS Collaboration \cite{PhysRevD.79.034503}, which employ the RG improved Iwasaki gauge action at $\beta=1.90$ and the non-perturbatively O(a)-improved Wilson quark action with $c_{\rm SW} = 1.715$ at $(\kappa_{\rm ud}, \kappa_{\rm s})=(0.13781, 0.13640)$.
This parameter set corresponds to the lattice spacing $a = 0.0907(14)~{\rm fm}$ (or $a^{-1} = 2.17(53)~{\rm GeV}$), the lattice spatial size $La = 2.902(42)~{\rm fm}$ and the pion mass $m_\pi \simeq 155.7~{\rm MeV}$.
Gauge configurations are fixed to both the Coulomb gauge and the Landau gauge.
The charm quark is introduced by the quenched approximation using the relativistic heavy quark (RHQ) action with the parameters in Ref.\cite{PhysRevD.84.074505}.

We calculate the two-point and four-point correlators of the $c\bar{c}$ system using quark propagators generated from a wall source and a point sink with the periodic boundary condition imposed on the temporal and the spatial directions.
Statistical data are binned with size $l_{\rm bin} = 11$.
Statistical precision is enhanced in several ways: (i) $64$ source points in the temporal direction are utilized through translational symmetry, and (ii) the propagators in both forward and backward directions are used through time reversal and
charge conjugation symmetries.

\subsection{Charmonium spectroscopy from two-point correlators}
\begin{figure}[t]
  \centering
  \includegraphics[width=0.4\textwidth]{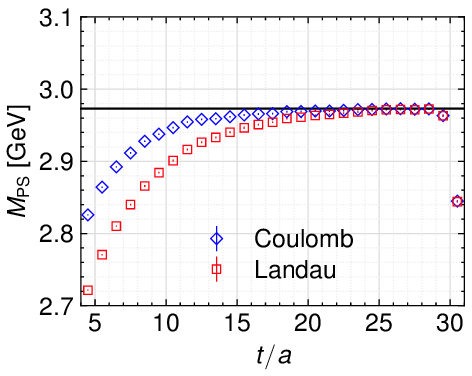}~\qquad\quad~\includegraphics[width=0.4\textwidth]{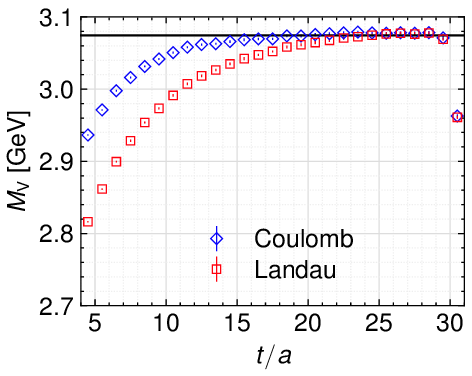}
  \caption{Effective mass plots of charmonium in pseudoscalar (PS, left) and vector (V, right) channels in Coulomb gauge (red) and Landau gauge (blue). The solid lines represent the masses extracted from single-cosh fits to the two-point correlators in Coulomb gauge.}
  \label{fig:effmass}
\end{figure}

The effective masses of charmonium calculated from the quenched lattice QCD simulation are shown in \Fig{fig:effmass}, where results in both Coulomb and Landau gauges are presented for the PS (corresponds to $\eta_c$) and V (corresponds to $J/\psi$) channels.
The effective mass is defined as
\begin{equation}\label{eq:effmass}
  M_\Gamma(t+\tfrac{a}{2}) = \ln\frac{C_\Gamma(t)}{C_\Gamma(t+a)},
\end{equation}
where $C_\Gamma(t)$ is the two-point correlator obtained by taking $\mathbf{r} = \mathbf{0}$ and $t_{\rm src} = 0$ in the four-point correlator $C_\Gamma(\mathbf{r}, t;t_{\rm src})$ defined in \Eq{eq:4pt},
\begin{equation}\label{eq:2pt}
  C_\Gamma(t) = \scaleobj{0.85}{\frac{1}{V}\sum_{\mathbf{\Delta}}\sum_{c}}\bigl\langle 0\bigl|T\bigl[\bar{q}_c(\mathbf{\Delta}, t)\Gamma q_c(\mathbf{\Delta}, t)\mathcal{J}_\Gamma^\dagger(0)\bigr]\bigr|0\bigr\rangle.
\end{equation}
The solid lines represent the masses extracted from single-cosh fits to the two-point correlators in Coulomb gauge.
As expected from the gauge independence of charmonium masses, the effective masses converge to the same value.
However, in both PS and V channels, the plateaus appear at smaller $t$ in the Coulomb gauge compared to the Landau gauge, indicating that the wall source operators in the Coulomb gauge have better overlap with the ground states.

A clear plateau is observed in the range $24.5 \leq t/a \leq 28.5$ for the Coulomb gauge, while the points beyond $t/a = 29.5$ show boundary artifacts.
To avoid these artifacts and account for the periodic boundary condition in the temporal direction, we fit the charmonium masses using a single-cosh form, obtaining $M_{\rm PS}=2.9729(16)~{\rm GeV}$ and $M_{\rm V}=3.0745(19)~{\rm GeV}$.

\subsection{Ground-state NBS wave functions from four-point correlators}
\begin{figure}[thp]
  \centering
  \begin{minipage}{0.48\textwidth}
    \includegraphics[width=0.49\linewidth]{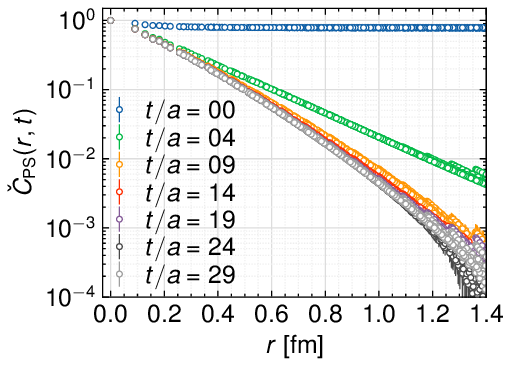}\hfill\includegraphics[width=0.49\linewidth]{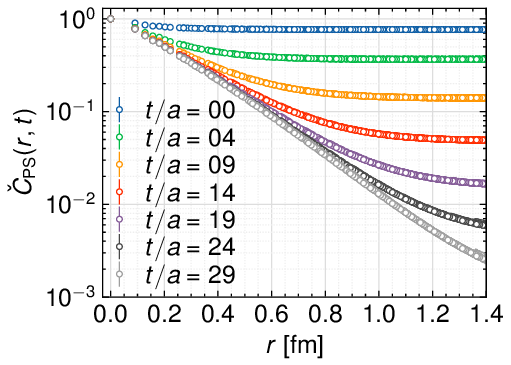}
    \caption{The normalized four-point correlation functions $\check{C}_\Gamma(\mathbf{r},t)$ in the PS channel at fixed time $t=0,4,9,14,19,24$ and $29$, shown for Coulomb gauge (left) and Landau gauge (right). $\check{C}_\Gamma(\mathbf{r},t)$ is plotted against $r\equiv|\mathbf{r}|$.}
    \label{fig:4-point}
  \end{minipage}
  \hfill
  \begin{minipage}{0.48\textwidth}
    \includegraphics[width=0.49\linewidth]{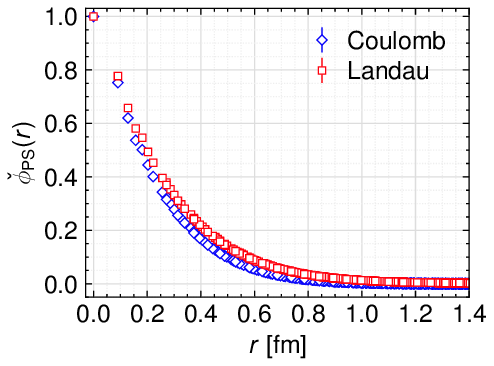}\hfill\includegraphics[width=0.49\linewidth]{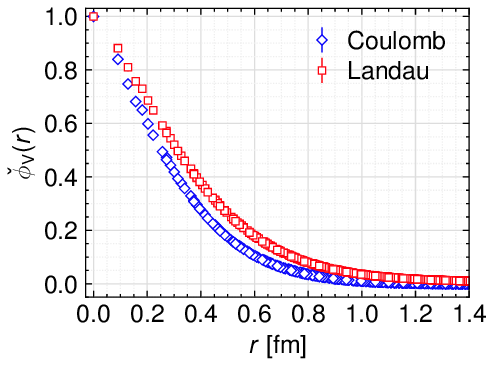}
    \caption{The normalized NBS wave functions in Coulomb gauge (blue) and Landau gauge (red), shown in PS (left) and V (right) channels. The NBS wave functions are normalized by $\check{\phi}_\Gamma(r) = \phi_\Gamma(r) / \phi_\Gamma(0)$.}
    \label{fig:NBSwave}
  \end{minipage}
\end{figure}

In the analysis of four-point correlators, we restrict our data to $t/a \leq 29$ to avoid the boundary artifacts.
The figures in this section display results as functions of the interquark distance $r\equiv|\mathbf{r}|$.
\Fig{fig:4-point} shows the normalized four-point correlators defined by
\begin{equation}
  \check{C}_\Gamma(\mathbf{r},t) = C_\Gamma(\mathbf{r},t) / C_\Gamma(0,t).
\end{equation}
The correlators in the Coulomb gauge exhibit quick convergence with respect to time.
In contrast, the Landau gauge correlators show slower convergence, particularly in the long-range region, where they show plateau-like behavior.
This behavior is consistent with our observations from the effective mass plots (\Fig{fig:effmass}), where the plateaus appear at smaller $t$ for the Coulomb gauge compared to the Landau gauge.
Within the spatial region $r\lesssim 0.65~{\rm fm}$, the correlators achieve approximate convergence at $t/a = 29$.
Similar behavior is observed in the V channel.

Hereafter, unless otherwise specified, we consider the four-point correlators at $t/a = 29$ to be dominated by the ground state contribution and interpret them as the NBS wave functions.
The comparison of the normalized ground-state NBS wave function in Coulomb gauge and Landau gauge is shown in \Fig{fig:NBSwave}.
The NBS wave functions are wider in Landau gauge than in Coulomb gauge both in PS and V channels.
A similar tendency is also reported for the quark-diquark NBS wave function for the nucleon in Ref.~\cite{PhysRevD.76.074021}.

\subsection[Charm quark masses and ccbar potentials]{Charm quark masses and $c\bar{c}$ potentials}
\begin{figure}[thp]
  \centering
  \includegraphics[width=0.45\textwidth]{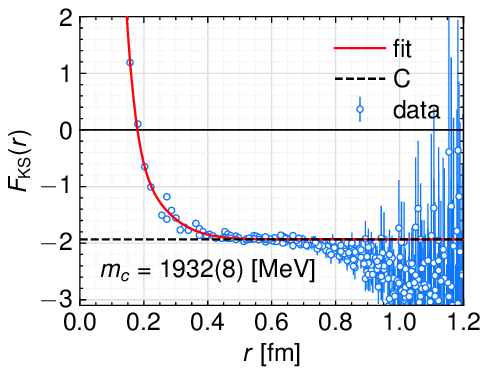}~\qquad~\includegraphics[width=0.45\textwidth]{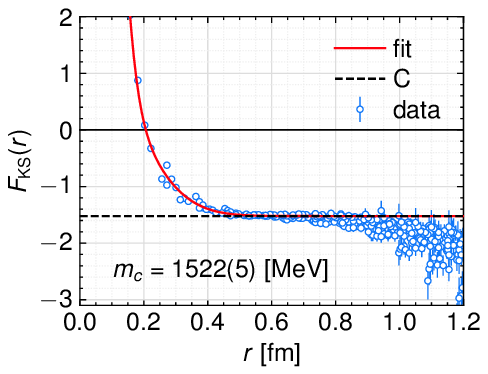}
  \caption{Kawanai-Sasaki function for Coulomb gauge (left) and Landau gauge (right). The red curves represent fits using the two-Gaussian functional form given in \Eq{eq:2gaussians}. The horizontal dashed lines indicate the asymptotic value $C$, which corresponds to the negative charm quark mass $-m_c$.}
  \label{fig:KSfunction}
\end{figure}

The charm quark masses are determined by the Kawanai-Sasaki condition \Eq{eq:KS_condition}.
Kawanai-Sasaki functions $F_{\rm KS}(\mathbf{r})$ for different gauges are shown in \Fig{fig:KSfunction}.
We fit the Kawanai-Sasaki functions using a two-Gaussian parametrization defined as
\begin{equation}\label{eq:2gaussians}
  f(r) \equiv \sum_{n=1}^{2} a_n \exp(-\nu_n r^2) + C,
\end{equation}
where $a_n, \nu_n(n=1,2)$ and $C$ are the fit parameters.
Here $C$ is responsible for the constant behavior at the long distance, so the charm quark mass is obtained as $m_c=-C$.
The fit results are also shown in \Fig{fig:KSfunction} denoted by red curves, through which we obtain $m_c = 1932(8)~{\rm MeV}$ for Coulomb gauge and smaller $m_c = 1522(5)~{\rm MeV}$ for Landau gauge.

\begin{figure}[thp]
  \centering
  \includegraphics[width=0.45\textwidth]{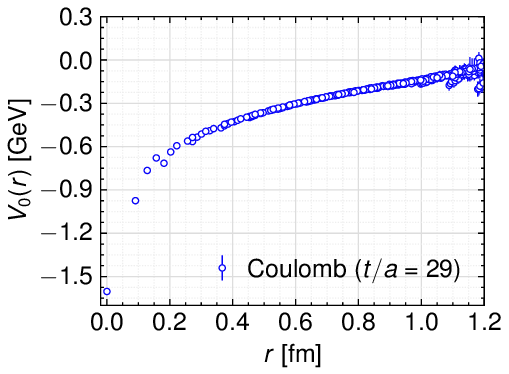}~\qquad~\includegraphics[width=0.45\textwidth]{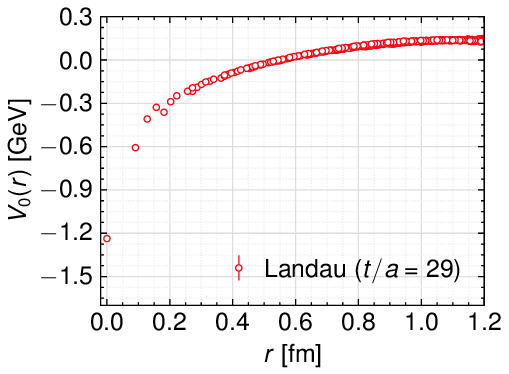}
  \caption{The central potential $V_0(r)$ for charmonium in Coulomb gauge (left) and Landau gauge (right).}
  \label{fig:V0separate}
\end{figure}

\begin{figure}[thp]
  \centering
  \includegraphics[width=0.45\textwidth]{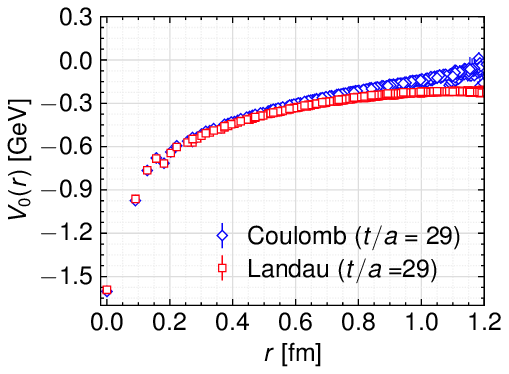}~\qquad~\includegraphics[width=0.45\textwidth]{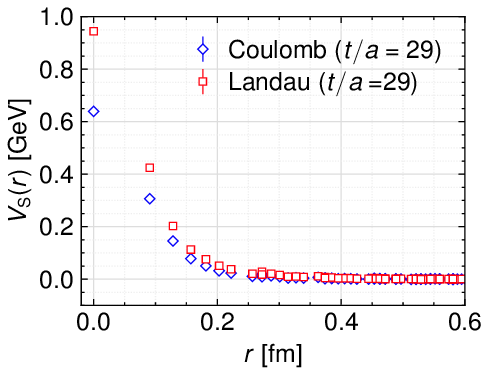}
  \caption{The central potential $V_0(r)$ (left) and the spin-dependent potential $V_{\rm S}(r)$ (right) for $c\bar{c}$ system in Coulomb gauge (blue) and Landau gauge (red). The central potential $V_0(r)$ in Landau gauge is vertically shifted for direct comparison.}
  \label{fig:V0Vs}
\end{figure}

The potentials are determined through the HAL QCD method, incorporating the calculated charmonium masses, charm quark masses, and ground-state NBS wave functions.
\Fig{fig:V0separate} presents individual plots of the central potentials $V_0(\mathbf{r})$ in both Coulomb and Landau gauges.
\Fig{fig:V0Vs} illustrates gauge dependence of the central potentials $V_0(\mathbf{r})$ (with Landau gauge results vertically offset for clarity) and the spin-dependent potentials $V_{\rm S}(\mathbf{r})$.

The central potential $V_0(\mathbf{r})$ in the Coulomb gauge (\Fig{fig:V0separate}, left panel) exhibits Cornell-type behavior throughout the entire spatial region.
In contrast, $V_0(\mathbf{r})$ in the Landau gauge (\Fig{fig:V0separate}, right panel) demonstrates Coulomb-like behavior at short distances but deviates from the expected linear-rising behavior at large distances.
The superposition of data from both gauges (\Fig{fig:V0Vs}, left panel) enables a direct comparison of their behaviors.
The potentials agree well in the region $r \lesssim 0.5~{\rm fm}$, beyond which their difference progressively increases.
The deviation becomes significant beyond $r \sim 0.6~{\rm fm}$, coinciding with the spatial region ($r \gtrsim 0.65~{\rm fm}$) where the four-point correlators show substantial excited-state contamination.
This correspondence suggests that the absence of linear-rising behavior at large distances in the Landau gauge potential stems from insufficient ground-state dominance in the four-point correlators.

The spin-dependent potentials $V_{\rm S}(\mathbf{r})$ (\Fig{fig:V0Vs}, right panel) can be characterized as vertically shifted $F_{\rm KS}(\mathbf{r})$ with a scaling factor of $1/m_c$.
These potentials are spatially localized within the region of $r \lesssim 0.4~{\rm fm}$.
The spatial extent of these potentials remains consistent between both Coulomb and Landau gauges.
Stronger $V_{\rm S}(\mathbf{r})$ is observed in the Landau gauge compared to that in the Coulomb gauge.

\subsection{Comments on the results}
If we assume that the central potentials $V_0(\mathbf{r})$ in both Coulomb and Landau gauges agree, the observed behaviors of charm quark masses, equal-time NBS wave functions, and spin-dependent potentials can be understood through non-relativistic perturbation theory. In this framework, the Hamiltonian in \Eq{eq:Seq_for_ccbar} is decomposed into the unperturbed term $H_0 = -\nabla^2/(2\mu) + V_0(\mathbf{r})$ and the perturbation term $V_{\rm S}(\mathbf{r})\mathbf{s}_1\cdot\mathbf {s}_2$.

The NBS wave functions obtained in this work exhibit broader spatial distributions in the Landau gauge than in the Coulomb gauge (\Fig{fig:NBSwave}).
This can be explained, at the leading order of perturbation theory, by the smaller charm quark mass $m_c$ in the Landau gauge compared to that in the Coulomb gauge.
At next-to-leading order, reproducing the gauge-invariant hyperfine mass splitting $M_{\rm V} - M_{\rm PS}$ with broader wave functions in the Landau gauge requires stronger spin-dependent potential $V_{\rm S}(\mathbf{r})$, as confirmed by our calculations (\Fig{fig:V0Vs}).

\section{Summary}

We investigated the gauge dependence of $c\bar{c}$ potentials and charm quark masses from the NBS amplitude method with the Kawanai-Sasaki prescription.
Our calculations were performed in both Coulomb and Landau gauges using 2+1 flavor QCD configurations.
The central potentials obtained in both gauges show good agreement at short distances ($r \lesssim 0.5$ fm), while exhibiting significant differences at larger distances.
In the Landau gauge, we observe that the expected linear-rising behavior of the central potential is suppressed at large distance, which we attribute to insufficient ground-state dominance in the four-point correlators.
Our analysis reveals three consistent features in the Landau gauge results: a broader NBS wave function, a smaller charm quark mass, and a stronger spin-dependent potential compared to the Coulomb gauge.
These characteristics are mutually consistent within the framework of non-relativistic perturbation theory.

\acknowledgments{The lattice QCD calculations in this work were performed on the supercomputer SQUID at the Cyber Media Center (CMC), Osaka University, under the support by the Research Center for Nuclear Physics (RCNP), Osaka University.
  We acknowledge the PACS-CS Collaboration for providing the 2+1 flavor QCD gauge configurations, and we thank the Japan Lattice Data Grid (JLDG) and International Lattice Data Grid (ILDG) for their data-sharing infrastructure.
  Our numerical calculations utilized a modified version of the lattice QCD library Bridge++ \cite{Ueda_2014}, whose original codebase is publicly available at \href{http://bridge.kek.jp/Lattice-code/}{http://bridge.kek.jp/Lattice-code/}.
  This work was supported by JSPS KAKENHI Grant Number JP21K03535.
  Additional support was provided by JST, the establishment of university fellowships towards the creation of science technology innovation, Grant Number JPMJFS2125.}

\clearpage
\bibliographystyle{JHEP}
\bibliography{bibdb}

\providecommand{\href}[2]{#2}\begingroup\raggedright\begin{thebibliography}{10}

\bibitem{PhysRevLett.99.022001}
N.~Ishii, S.~Aoki and T.~Hatsuda, \emph{Nuclear force from lattice {QCD}}, \href{https://doi.org/10.1103/PhysRevLett.99.022001}{\emph{Phys. Rev. Lett.} {\bfseries 99} (2007) 022001} [\href{https://arxiv.org/abs/nucl-th/0611096}{{\ttfamily nucl-th/0611096}}].

\bibitem{10.1143/PTP.123.89}
S.~Aoki, T.~Hatsuda and N.~Ishii, \emph{Theoretical foundation of the nuclear force in {QCD} and its applications to central and tensor forces in quenched lattice {QCD} simulations}, \href{https://doi.org/10.1143/PTP.123.89}{\emph{Prog. Theor. Phys.} {\bfseries 123} (2010) 89} [\href{https://arxiv.org/abs/0909.5585}{{\ttfamily 0909.5585}}].

\bibitem{ISHII2012437}
N.~Ishii, S.~Aoki, T.~Doi, T.~Hatsuda, Y.~Ikeda, T.~Inoue et~al., \emph{Hadron-hadron interactions from imaginary-time {Nambu-Bethe-Salpeter} wave function on the lattice}, \href{https://doi.org/10.1016/j.physletb.2012.04.076}{\emph{Phys. Lett. B} {\bfseries 712} (2012) 437} [\href{https://arxiv.org/abs/1203.3642}{{\ttfamily 1203.3642}}].

\bibitem{PhysRevLett.107.091601}
T.~Kawanai and S.~Sasaki, \emph{Interquark potential with finite quark mass from lattice {{QCD}}}, \href{https://doi.org/10.1103/PhysRevLett.107.091601}{\emph{Phys. Rev. Lett.} {\bfseries 107} (2011) 091601} [\href{https://arxiv.org/abs/1102.3246}{{\ttfamily 1102.3246}}].

\bibitem{PhysRevD.85.091503}
T.~Kawanai and S.~Sasaki, \emph{Charmonium potential from full lattice {{QCD}}}, \href{https://doi.org/10.1103/PhysRevD.85.091503}{\emph{Phys. Rev. D} {\bfseries 85} (2012) 091503} [\href{https://arxiv.org/abs/1110.0888}{{\ttfamily 1110.0888}}].

\bibitem{PhysRevD.89.054507}
T.~Kawanai and S.~Sasaki, \emph{Heavy quarkonium potential from {Bethe-Salpeter} wave function on the lattice}, \href{https://doi.org/10.1103/PhysRevD.89.054507}{\emph{Phys. Rev. D} {\bfseries 89} (2014) 054507} [\href{https://arxiv.org/abs/1311.1253}{{\ttfamily 1311.1253}}].

\bibitem{PhysRevD.92.094503}
T.~Kawanai and S.~Sasaki, \emph{Potential description of charmonium and charmed-strange mesons from lattice {{QCD}}}, \href{https://doi.org/10.1103/PhysRevD.92.094503}{\emph{Phys. Rev. D} {\bfseries 92} (2015) 094503} [\href{https://arxiv.org/abs/1508.02178}{{\ttfamily 1508.02178}}].

\bibitem{PhysRevD.94.114514}
K.~Nochi, T.~Kawanai and S.~Sasaki, \emph{{Bethe-Salpeter} wave functions of ${\ensuremath{\eta}}_{c}(2s)$ and $\ensuremath{\psi}(2s)$ states from full lattice {QCD}}, \href{https://doi.org/10.1103/PhysRevD.94.114514}{\emph{Phys. Rev. D} {\bfseries 94} (2016) 114514} [\href{https://arxiv.org/abs/1608.02340}{{\ttfamily 1608.02340}}].

\bibitem{Watanabe2021}
K.~Watanabe and N.~Ishii, \emph{Building diquark model from lattice {QCD}}, \href{https://doi.org/10.1007/s00601-021-01627-y}{\emph{Few-Body Systems} {\bfseries 62} (2021) 45} [\href{https://arxiv.org/abs/2105.07969}{{\ttfamily 2105.07969}}].

\bibitem{PhysRevD.105.074510}
K.~Watanabe, \emph{Quark-diquark potential and diquark mass from lattice {QCD}}, \href{https://doi.org/10.1103/PhysRevD.105.074510}{\emph{Phys. Rev. D} {\bfseries 105} (2022) 074510} [\href{https://arxiv.org/abs/2111.15167}{{\ttfamily 2111.15167}}].

\bibitem{Ikeda:2011mW}
Y.~Ikeda, \emph{The $\bar{q} q$ potential from {Bethe - Salpeter} amplitudes on lattice}, \href{https://doi.org/10.22323/1.105.0143}{\emph{PoS} {\bfseries LATTICE2010} (2011) 143}.

\bibitem{Iida:20123l}
H.~Iida and Y.~Ikeda, \emph{Inter-quark potentials from {NBS} amplitudes and their applications}, \href{https://doi.org/10.22323/1.139.0195}{\emph{PoS} {\bfseries LATTICE2011} (2012) 195}.

\bibitem{10.1143/PTP.128.941}
Y.~Ikeda and H.~Iida, \emph{Quark-anti-quark potentials from {Nambu-Bethe-Salpeter} amplitudes on lattice}, \href{https://doi.org/10.1143/PTP.128.941}{\emph{Prog. Theor. Phys.} {\bfseries 128} (2012) 941} [\href{https://arxiv.org/abs/1102.2097}{{\ttfamily 1102.2097}}].

\bibitem{PhysRevD.79.034503}
{\scshape PACS-CS} collaboration, \emph{$2+1$ flavor lattice {QCD} toward the physical point}, \href{https://doi.org/10.1103/PhysRevD.79.034503}{\emph{Phys. Rev. D} {\bfseries 79} (2009) 034503} [\href{https://arxiv.org/abs/0807.1661}{{\ttfamily 0807.1661}}].

\bibitem{PhysRevD.84.074505}
{\scshape PACS-CS} collaboration, \emph{Charm quark system at the physical point of $2\mathbf{+}1$ flavor lattice {QCD}}, \href{https://doi.org/10.1103/PhysRevD.84.074505}{\emph{Phys. Rev. D} {\bfseries 84} (2011) 074505} [\href{https://arxiv.org/abs/1104.4600}{{\ttfamily 1104.4600}}].

\bibitem{PhysRevD.76.074021}
R.~Babich, N.~Garron, C.~Hoelbling, J.~Howard, L.~Lellouch and C.~Rebbi, \emph{Diquark correlations in baryons on the lattice with overlap quarks}, \href{https://doi.org/10.1103/PhysRevD.76.074021}{\emph{Phys. Rev. D} {\bfseries 76} (2007) 074021} [\href{https://arxiv.org/abs/hep-lat/0701023}{{\ttfamily hep-lat/0701023}}].

\bibitem{Ueda_2014}
S.~Ueda, S.~Aoki, T.~Aoyama, K.~Kanaya, H.~Matsufuru, S.~Motoki et~al., \emph{Development of an object oriented lattice {QCD} code "bridge++"}, \href{https://doi.org/10.1088/1742-6596/523/1/012046}{\emph{J. Phys. Conf. Ser.} {\bfseries 523} (2014) 012046}.

\end{thebibliography}\endgroup

\end{document}